\documentclass{svjour3}
\usepackage{mathptmx}
\usepackage{amsmath, amsbsy, amsfonts}
\usepackage{epsfig}
\newcommand{\gr}{\mbox{$\nabla$}}

\newcommand{\dv}{\mbox{$\mbox{div}$}}

\newcommand{\Kb}{\mathbb{K}}
\newcommand{\xb}{\mbox{\bf x}}

\newcommand{\der}[2]{\frac{\partial #1}{\partial #2}}
\newcommand{\dert}[1]{\frac{\partial #1}{\partial t}}

\newcommand{\ql}{\mbox{$\mathbf{q}_l$}}
\newcommand{\qg}{\mbox{$\mathbf{q}_g$}}
\newcommand{\gb}{\mbox{$\mathbf{g}$}}
\newcommand{\jlw}{\mbox{$\mathbf{j}_l^w$}}
\newcommand{\jlh}{\mbox{$\mathbf{j}_l^h$}}
\newcommand{\jgw}{\mbox{$\mathbf{j}_g^w$}}
\newcommand{\jgh}{\mbox{$\mathbf{j}_g^h$}}
\newcommand{\Jb}{\mbox{$\mathbf{J}$}}
\newcommand{\jalphai}{\mbox{$\mathbf{j}_{\alpha}^i$}}
\newcommand{\jli}{\mbox{$\mathbf{j}_l^i$}}
\newcommand{\jgi}{\mbox{$\mathbf{j}_g^i$}}
\newcommand{\bphi}{\boldsymbol{\phi}}
\newcommand{\Htwo}{$\text{H}_2$}
\author{Alain Bourgeat \and Mladen Jurak \and  Farid Sma\"{\i}}
\institute{Alain Bourgeat \at Universit\'{e} de Lyon, Universit\'{e} Lyon1,
CNRS UMR 5208 Institut Camille Jordan, F - 69200 Villeurbanne Cedex,
France
    \and Mladen Jurak \at Department of mathematics, University of Zagreb, Croatia,
    \and Farid Sma\"{\i} \at Universit\'{e} de Lyon, Universit\'{e} Lyon1,
CNRS UMR 5208 Institut Camille Jordan, F - 69200 Villeurbanne Cedex,
France}
\date{\today}
\title{Two phase partially miscible flow and transport modeling in porous media ; application to gas migration in
 a nuclear waste repository}
\begin{document}
\maketitle

\begin{abstract}
We derive a compositional compressible two-phase, liquid and gas,
flow model for numerical simulations of hydrogen migration in deep
geological repository for radioactive waste. This model includes
capillary effects and the gas high diffusivity. Moreover, it is
written in variables (total hydrogen mass density and liquid
pressure) chosen in order to be consistent with gas appearance or
disappearance. We discuss the well possedness of this model and give
some computational evidences of its adequacy to simulate gas
generation in a water saturated repository.
\end{abstract}
 \tableofcontents

\section{Introduction}

The simultaneous flow of immiscible fluids in porous media occurs in
a wide variety of applications.
The most concentrated research in the field of multiphase flows over
the past four decades has focused on  unsaturated groundwater flows,
and flows in underground petroleum reservoirs. Most recently,
multiphase flows have generated serious interest among engineers
concerned with  deep geological repository for radioactive waste.
There is growing awareness that the effect of hydrogen gas
generation, due to anaerobic corrosion of the steel engineered
barriers (carbon steel overpack and stainless steel envelope) of
radioactive waste packages, can affect all the functions allocated
to the canisters, waste forms, buffers, backfill, host rock. Host
rock safety function may be threaten by overpressurisation leading
to opening fractures of the host rock, inducing groundwater flow and
transport of radionuclides.

 The equations governing these flows are inherently nonlinear,
and the geometries and material properties characterizing many
problems in many applications can be quite irregular and contrasted.
As a result, numerical simulation often offers the only viable
approach to the mathematical modeling of multiphase flows. In nuclear
waste management, the migration of gas through the near field
environment and the host rock,
 involves two components, water and pure hydrogen \Htwo; and
two phases "liquid" and "gas". It is then not clear if conventional
models, like for instance the Black-Oil model,  used in petroleum or
groundwater engineering are still valid  for such a situation. Our
ability to understand and predict  underground gas migration  is
crucial to the design of reliable waste storages. This is a fairly
new frontier in multiphase porous-media flows, and again the
inherent complexity of the physics leads to governing equations for
which the only practical way to produce solutions may be numerical
simulation.
 This exposition provides an overview of the types of standard models
that are used in the field of compressible multiphase multispecies
flows in porous media and includes discussions  of the problems
coming from  \Htwo\  gas being one of the  components.
 Finally the paper also addresses one of the outstanding physical and
 mathematical problems in
multiphase flow simulation:  the appearance disappearance of one of
the phases, leading to the degeneracy of the equations satisfied by
the saturation. It has been seen recently in a benchmark organized
by the French agency in charge of the Nuclear Waste management in
France \cite{Bench} that none of the usual codes used in that field
where able to simulate adequately the appearance or/and
disappearance of one of the phases. In order to overcome this
difficulty, we will  discuss a formulation
 based on new variables  which doesn't degenerate.
We will demonstrate through two test cases, the ability of this new
formulation to actually cope with the appearance or/and
disappearance  of one phase. The scope of the paper is limited to
isothermal flows in rigid porous media and will not include the
possible process of "pathway dilation" as described in
\cite{Hors}.

\section{Conceptual and mathematical model}
 Our goal is first to present a survey of the conventional models used for describing
 two-phase two-components flow in porous media. Most of  these models have been
 designed and widely used in petroleum engineering (see for
 instance: \cite{Allen}, \cite{Bear:1}, \cite{ChJa}, \cite{DPeace}, \cite{ECLIPSE}).
We consider herein a
 porous medium saturated with a fluid composed of 2 phases :
 \textit{liquid} and \textit{gas}. According to the application we have in mind,
  the fluid is a mixture of two
 components: water (mostly liquid)  and  hydrogen (\Htwo, mostly gas).
 Water is present in the gas phase through vaporization, and hydrogen
 is present in the liquid phase through dissolution. The fluids are
 compressible and the model is assumed to be isothermal. For
 simplicity, we assume first the porous medium to be rigid, meaning  the
 porosity $\Phi$ is only a function of the space variable
 $\Phi=\Phi(\xb)$; and second, we neglect the pressure induced dilation of
 gas pathways.
 Hydrogen being  highly  diffusive  we have described in detail how the diffusion
 could be taken in account in the models.

\subsection{Petrographic and fluid properties}
 \subsubsection{Fluid phases}
The two phases will be denoted by indices $l$, for liquid and $g$
for gas. Associated to each phase in the porous media are the following quantities:
\begin{itemize}
\item  liquid and gas phase pressures: $p_l$, $p_g$;
\item  liquid and gas phase saturations: $S_l$, $S_g$;
\item  liquid and gas phase mass densities: $\rho_l$, $\rho_g$;
\item  liquid and gas phase viscosities: $\mu_l$, $\mu_g$;
\item phase volumetric flow rates, \ql\ and \qg;
\end{itemize}
Then,
 the {\em  Darcy-Muskat law} says:
\begin{equation}
 \ql = - \Kb(\xb)\frac{kr_l(S_l)}{\mu_l}
          \left( \gr p_l -\rho_l \gb\right),\quad
          \qg  = - \Kb(\xb)\frac{kr_g(S_g)}{\mu_g}
          \left( \gr p_g -\rho_g \gb\right),\label{darcy}
\end{equation}
where $ \Kb(\xb)$ is the absolute permeability tensor, $kr_l$ and
$kr_g$ are the relative permeability functions, and $\gb$ is the
gravity acceleration.

The above phase mass densities and viscosities are all functions of
phase pressures and of  phase {\em composition}. Moreover, phase
saturations satisfy
\begin{equation}
S_l + S_g = 1     \label{saturations}
\end{equation}
and the pressures are connected through a given experimental  {\em
capillary pressure law}:
\begin{equation}
 p_c(S_g) = p_g -p_l.\label{capillary}
\end{equation}
From definition (\ref{capillary}) we should notice that $p_c$ is strictly
increasing function of gas saturation, $p_c'(S_g) > 0$.

\subsubsection{Fluid components}

Water and pure hydrogen components will be denoted by indices $w$
and $h$. Since the liquid phase could be composed of water and
dissolved hydrogen we need to introduce the water mass density in
the liquid phase $\rho_l^w$, and the hydrogen mass density in the
liquid phase $\rho_l^h$. Similarly, we introduce the water and
hydrogen mass densities in the gas phase: $\rho_g^w$, $\rho_g^h$.
Note that the upper index is always the component  index, and the lower one
denotes the phase. We have, then
\begin{equation}
\rho_l = \rho_l^w + \rho_l^h,\quad   \rho_g = \rho_g^w + \rho_g^h .
\label{partial-densities}
\end{equation}
Since the composition of each phase is generally unknown we
introduce {\em the mass fraction}  of the component
$i\in\{w,h\}$ in the phase  $\alpha\in\{g,l\}$
\begin{equation}
\omega_\alpha^i = \frac{\rho_\alpha^i}{\rho_\alpha};\quad
\omega_\alpha^w +\omega_\alpha^h =1,\;\alpha\in\{g,l\} .
\label{fractions}
\end{equation}

\subsection{Mass conservation equation}
The conservation of the mass applies to each component $i$ and
reads, for any arbitrary control volume  ${\cal R}$:
\begin{align}
\frac{dm^i}{dt} + F^i = {\cal F}^i \;, \;\;i\in\{w,h\}. \label{ZSM}
\end{align}
where
\begin{itemize}
\item $m^i$ is the mass of the  component $i$, in the control volume ${\cal R}$,
    at given instant $t$;
\item $F^i$ is the rate at which the  component $i$ is leaving (migrating from)
    the  volume ${\cal R}$, at given instant $t$;
\item ${\cal F}^i$ is the source term (the rate at which the $i$ component is added
    to ${\cal R}$ by the source).
\end{itemize}
The mass of each component $i$ is a sum over all phases $g$ and $l$:
\begin{align*}
 m^i  = \int_{\cal R} \Phi\left(  S_l \rho_l \omega_l^i + S_g \rho_g\omega_g^i \right)\,
 d\xb, \quad i\in\{w,h\}.
\end{align*}
In each phase, the migration of a component is due to the transport
by the phase velocity and to  the molecular diffusion:
\begin{align*}
F^i &= \int_{\partial {\cal R}} \left(  \rho_l \omega_l^i  \ql +
\rho_g \omega_g^i   \qg  +\jli +\jgi\right) \cdot {\bf n}\,
d\xb;\;\;i\in\{w,h\};
\end{align*}
where ${\bf n}$ is the unit outer normal to ${\partial {\cal R}}$.
 The phase
flow velocities, $\ql$ and $\qg$ are given by the Darcy-Muskat
law (\ref{darcy}), and the component $i$  diffusive flux
in  phase $\alpha$ is denoted
 $\jalphai$, $i\in\{w,h\}$, $\alpha\in\{g,l\}$, and will be defined in the next
paragraph, by equations (\ref{diff-fluxes}).
From (\ref{ZSM}) we get the differential equations:
\begin{align}
 \Phi\dert{}\left(  S_l \rho_l \omega_l^w + S_g \rho_g\omega_g^w \right)   +
\dv \left(  \rho_l \omega_l^w  \ql + \rho_g \omega_g^w   \qg   +\jlw +\jgw\right)
= {\cal F}^w, \label{eq-1} \\
\Phi\dert{}\left(  S_l \rho_l\omega_l^h  + S_g  \rho_g\omega_g^h\right) +
\dv \left(  \rho_l \omega_l^h \ql +  \rho_g \omega_g^h \qg +\jlh +\jgh\right)
= {\cal F}^h .\label{eq-2}
\end{align}

\subsubsection{Diffusion fluxes}

From $M^w$ and $M^h$, the water and hydrogen  molar masses, using
definitions (\ref{fractions}) we define the following water and
hydrogen molar concentrations in each phase $\alpha\in\{g,l\}$:
\begin{equation}
c_\alpha^h = \frac{S_\alpha \rho_\alpha^h}{M^h} =  \frac{S_\alpha
\rho_\alpha \omega_\alpha^h}{M^h},\quad c_\alpha^w = \frac{S_\alpha
\rho_\alpha^w}{M^w} = \frac{S_\alpha \rho_\alpha
\omega_\alpha^w}{M^w}. \label{molar-concentrations}
\end{equation}
Then the phase $\alpha$ molar concentrations,  $\alpha\in\{g,l\}$,
is:
\begin{equation}
 c_\alpha = c_\alpha^h + c_\alpha^w = S_\alpha \rho_\alpha \left( \frac{\omega_\alpha^h}{M^h} +
\frac{\omega_\alpha^w}{M^w}\right).\label{phase-molar-concentrations}
\end{equation}
Usually,  component $i$ diffusive flux in phase $\alpha$ is assumed
to be depending on $X_\alpha^i$, the component $i$ molar fraction in
 phase $\alpha\in\{g,l\}$, defined from
(\ref{molar-concentrations}) and (\ref{phase-molar-concentrations}):
\begin{equation}
  X_\alpha^h = \frac{c_\alpha^h}{c_\alpha}
  = \frac{\omega_\alpha^h}{\omega_\alpha^h +(M^h/M^w)\omega_\alpha^w},\quad
  X_\alpha^w = \frac{c_\alpha^w}{c_\alpha}
  = \frac{\omega_\alpha^w}{\omega_\alpha^w +(M^w/M^h)\omega_\alpha^h};\quad
  \sum_{i=h,w} X_{\alpha}^i=1,
\label{molarfrac1}
\end {equation}
for $\alpha\in\{l,g\}$. Molar diffusive flux of component $i$ in  phase
$\alpha\in\{g,l\}$ is given by
 \[
   \mathbf{J}_\alpha^i  = - c_\alpha D_\alpha^i \gr X_\alpha^i;\quad
   \alpha\in\{g,l\}, \quad i\in\{w,h\}
   \]
and give molar diffusive flow rate of component $i$ through the unit
area. Coefficients $ D_\alpha^h$ and $ D_\alpha^w$ (unit $L^2/T$)
are Darcy scale molecular diffusion coefficients of components in phase
$\alpha\in\{g,l\}$. Mass flux of component $i$ in
phase $\alpha$, in equations (\ref{eq-1}), (\ref{eq-2}) is then
 obtained by multiplying the above component
  molar diffusive fluxes, $\mathbf{J}_\alpha^i $,
 by
the molar mass $M_i$ of the component $i $, and by the rock
porosity:
\begin{equation}
\mathbf{j}_\alpha^h  = -\Phi M^h c_\alpha  D_\alpha^h \gr
X_\alpha^h,\quad \mathbf{j}_\alpha^w  = -\Phi M^w c_\alpha
D_\alpha^w \gr X_\alpha^w. \label{diff-fluxes}
\end{equation}
\begin {remark}:
 Number of unknowns in the system is eight:
$ S_l, \rho_l^w,  \rho_l^h, p_l, S_g, \rho_g^w,  \rho_g^h, p_g$;
but, up to now, we have only  four equations (\ref{saturations}),
(\ref{capillary}), (\ref{eq-1}) and (\ref{eq-2}), and therefore,
four additional equations are needed to close the system.
\end {remark}
\begin {remark}:  Note  that $D_\alpha^h$ and $D_\alpha^w$ are not
exactly  molecular diffusion coefficients in phase $\alpha $,
corresponding to molecule-molecule interactions in free space but
$\Phi D_\alpha^i$, $\alpha\in\{g,l\}$, $i\in\{w,h\}$, are effective
diffusion coefficients, obtained
from the strict molecular diffusion coefficients by  a kind of
{\em averaging} through the
whole porous medium (see section 2.6 in \cite{Bear:1}, or
\cite{Milli} or \cite{Jin}). Moreover, for simplicity, we have not
included in diffusion of component $i$ in phase $\alpha $ any
dependancy on phase saturations; and then did not consider possible
non linear effects coming from coupling between advective-diffusive
transport ("dusty gas" model) and molecular streaming effects
(Knutsen diffusion) (see for instance \cite{Tough}).
\end {remark}
\begin{remark}
    In a binary system diffusive fluxes satisfy
    $\mathbf{j}_\alpha^h +\mathbf{j}_\alpha^w=0$,
    for $\alpha\in\{g,l\}$, and therefore we have
    \begin{equation}
        M^h D^h_\alpha = M^w D^w_\alpha,\quad \alpha\in\{g,l\}.
        \label{diff-coef-eq}
    \end{equation}
   \label{rem3}
\end{remark}

\subsection{Phase equilibrium Black-oil model}
\label{black-oil}

The additional equations needed to close the system of equations
(\ref{saturations}), (\ref{capillary}), (\ref{eq-1}) and
(\ref{eq-2}) will come from the assumption that the  two
phases are in equilibrium; equilibrium  meaning that at any  time the
quantity of hydrogen dissolved in the water is maximal for the given
pressure, and similarly, the  quantity of evaporated
water is maximal for the given pressure.
Composition of each phase is then uniquely determined by
its phase pressure  and saturation.
In this flow situation we say that water and gas phases are {\em saturated}.
 Nevertheless it could happen that one of the
phases disappears: either water can completely evaporate or hydrogen
can be completely dissolved in the water. Then in these situations
the composition of the remaining phase is not uniquely determined by
its phase pressure, and the phase composition becomes an independent
variable (instead of the saturation which is now constant, 0 or 1).
This flow situation correspond to the so-called {\em unsaturated
flow}. For unsaturated flow,  standard practice in petroleum
reservoir engineering is to introduce the following quantities:
\begin{itemize}
\item liquid and gas
{\em formation volume factors}, $B_l=B_l(p_l)$, $B_g=B_g(p_g)$;
\item {\em solution gas/liquid phase Ratio} $R_s=R_s(p_l)$;
\item {\em vapor water/gas phase Ratio} $R_v=R_v(p_g)$.
\end{itemize}
The explanation of formation volume factors and solution
component/phase ratios, as used in the oil reservoir modelling, is
as follows. Considering the volume $\Delta V_l^{res}$   of  liquid
at reservoir conditions (reservoir temperature and pressure); when
this volume of liquid is transported through the tubing to the
surface it separates   at  standard conditions to a volume  of
liquid $\Delta V_l^{std}$, and
a volume of gas $\Delta V_g^{std}$ (coming out of the liquid,
due to the pressure drop).
At standard (i.e. stock tank) conditions, the liquid phase contains
only oil component (here only water component)
    and the gas phase contains only gas component (here only hydrogen).
Then applying conservation of mass,
\[
\Delta V_l^{res}\rho_{l}= \Delta V_l^{std}\rho_l^{std} + \Delta
V_g^{std}\rho_{g}^{std};  \]
and denoting the solution gas/liquid
Ratio $R_{s}= {\Delta V_g^{std}}/{\Delta V_l^{std}}$, we may
write:
\[
\Delta V_l^{res}\rho_{l}= \Delta V_l^{std}(\rho_l^{std}
+R_{s}\rho_{g}^{std}),
   \]
and the liquid phase mass density decomposition
\begin{equation}
\rho_l=  \frac{\rho_l^{std} + R_s\rho_{g}^{std}}{B_l},
\label{density-petrol}
\end{equation}
where $B_l$ is the liquid formation volume factor,
      $B_l =   {\Delta V_l^{res}}/{\Delta V_l^{std}}$.
Similarly, from $\Delta V_g^{res}\rho_{g}$ and the gas formation
volume factor $B_{g}= {\Delta V_g^{res}}/{\Delta V_g^{std}}$, we
get the gas phase mass density decomposition
\begin{equation}
 \rho_g=  \frac{\rho_g^{std}+ R_v \rho_{l}^{std}}{B_g};
\label{density-petrol-gas}
\end{equation}
where the vapor water/gas phase ratio $R_{v}= {\Delta V_l^{std}}/{\Delta V_g^{std}}$.
 Now, in order to express the diffusion fluxes in terms of the new
variables $R_s, R_v, B_l$ and $B_g$ we have to rewrite the  molar concentrations
defined in (\ref{molar-concentrations}):
\begin{equation}
    c_l^h =  \frac{S_l R_s\rho_{g}^{std}}{M^h B_l}, \quad
    c_l^w =  \frac{S_l \rho_{l}^{std}}{M^w B_l},\quad
    c_g^h  =  \frac{S_g \rho_{g}^{std}}{M^h B_g},\quad
    c_g^w  =  \frac{S_g R_v\rho_{l}^{std}}{M^w B_g}.
\label{mol-con}
\end{equation}
These concentrations, defined in (\ref{mol-con}) may be slightly
different from those previously defined in
(\ref{molar-concentrations}) if the gas at
standard conditions is not composed only of hydrogen, and the liquid,
also at standard conditions, is note made only of water. The phase molar
concentrations, from (\ref{mol-con}), are then given by:
\[ c_l = c_l^h + c_l^w = \frac{S_l}{B_l}
\left( \frac{R_s\rho_{g}^{std}}{M^h} +  \frac{\rho_{l}^{std}}{M^w}\right)
= \frac{S_l}{B_l}  \frac{\rho_{g}^{std}}{M^h} (R_s+F) ,
\]
\[ c_g = c_g^h + c_g^w = \frac{S_g}{B_g}
\left( \frac{\rho_{g}^{std}}{M^h} +  \frac{R_v\rho_{l}^{std}}{M^w}\right)
= \frac{S_g}{B_g}  \frac{\rho_{g}^{std}}{M^h} (1+F  R_v) ,
\]
where $F$ is given by:
\begin{equation}
 F = \frac{M^h \rho_{l}^{std}}{M^w \rho_{g}^{std}}.
\label{F}
\end{equation}
The component {\em molar fractions} in phase, are now defined as:
\begin{equation}
\begin{aligned}
  X_l^h &=  \frac{c_l^h}{c_l}
  =  \frac{R_s\rho_{g}^{std}}{R_s\rho_{g}^{std} +(M^h/M^w)\rho_{l}^{std}} = \frac{R_s}{R_s+F},\\
  X_l^w &=  \frac{c_l^w}{c_l}
  =  \frac{\rho_l^{std}}{\rho_l^{std} +(M^w/M^h)R_s \rho_{g}^{std}} = \frac{F}{R_s+F},\\
  X_g^h &=  \frac{c_g^h}{c_g}
  =  \frac{\rho_{g}^{std}}{\rho_{g}^{std} +(M^h/M^w) R_v\rho_l^{std}} = \frac{1}{1+F R_v},\\
  X_g^w &=  \frac{c_g^w}{c_g}
  =  \frac{R_v\rho_{l}^{std}}{R_v\rho_{l}^{std} +(M^w/M^h)\rho_{g}^{std}} = \frac{F R_v}{1+F
  R_v}.
  \end{aligned}
  \label{molarfrac2}
\end{equation}
Mass diffusive fluxes of components, defined in
(\ref{diff-fluxes}),  depend on component {\em molar fractions} in
phases; they have then to be rewritten. For instance, the mass
diffusion flux of hydrogen in water, takes now the form:
\begin{align*}
\jlh/\rho_{g}^{std} &= -\Phi\frac{M^h}{\rho_{g}^{std}}\frac{S_l}{B_l}  \frac{\rho_{g}^{std}}{M^h}
\left(R_s + F \right)  D_l^h \gr X_l^h
 = -\Phi\frac{S_l}{B_l} \frac{F}{R_s + F} D_l^h\gr R_s.
\end{align*}
Other diffusion fluxes could be obtained by similar calculations,
leading to the following formulas:
\begin{align}
\bphi_{l}^{h}=
       \jlh/\rho_{g}^{std} &=  -\Phi\frac{S_l}{B_l} \frac{F}{R_s + F} D_l^h\gr R_s,\quad
\bphi_{l}^{w} = \jlw/\rho_{l}^{std} =  \Phi\frac{S_l}{B_l} \frac{1}{R_s + F} D_l^w\gr R_s,\\
\bphi_g^h = \jgh/\rho_{g}^{std} &=  \Phi\frac{S_g}{B_g} \frac{F}{1+F R_v} D_g^h\gr R_v,\quad
\bphi_g^w = \jgw/\rho_{l}^{std} =  -\Phi\frac{S_g}{B_g} \frac{1}{1+F R_v} D_g^w\gr R_v.
\end{align}
\begin{remark}
    Let us recall, like in Remark \ref{rem3}, that the diffusion coefficients  satisfy
    $M^h D_l^h = M^w D_l^w$, $M^h D_g^h = M^w D_g^w$; and, like
    in (\ref{molarfrac1}), $\sum_{i=h,w} X^i_{\alpha}=1,$
     for any phase
    $\alpha\in \{l,g\}$, in (\ref{molarfrac2}).
\end{remark}
The Darcy fluxes (\ref{darcy}) can now be rewritten in the following form:
\begin{align*}
\ql = -   \Kb\frac{kr_l}{\mu_l}
          \left( \gr p_l -\frac{\rho_{l}^{std} + R_s\rho_{g}^{std}}{B_l} \gb\right),\quad
\qg = -  \Kb\frac{kr_g}{\mu_g}
          \left( \gr p_g -\frac{\rho_{g}^{std}+ R_v \rho_{l}^{std}}{B_g} \gb\right),
\end{align*}
and the component fluxes, normalized by standard densities, are,
\begin{align}
\boldsymbol{\phi}^w
=  \frac{1}{B_l}\ql + \frac{R_v}{B_g} \qg + \bphi_{l}^{w} + \bphi_{g}^{w},\quad
\boldsymbol{\phi}^h
= \frac{R_s}{B_l}\ql +\frac{1}{B_g} \qg  + \bphi_{l}^{h} + \bphi_{g}^{h}.
\label{comp:flux}
\end{align}
Finally, we can write mass conservation (\ref{eq-1}), (\ref{eq-2}) in the following form:
\begin{align}
\Phi\dert{}\left( \frac{S_l}{B_l}  +\frac{R_v S_g}{B_g}  \right)  +
\dv &\left( \boldsymbol{\phi}^w
                  \right)={\cal F}^w/\rho_{l}^{std},  \label{main-eq-1}\\
\Phi\dert{}\left( \frac{  S_l R_s}{B_l}   + \frac{ S_g }{B_g}\right) +
\dv &\left(\boldsymbol{\phi}^h\right)
= {\cal F}^h/\rho_{g}^{std}. \label{main-eq-2}
\end{align}
These equations have to be completed by equations
(\ref{saturations}) and (\ref{capillary}). For instance, in
(\ref{main-eq-1}) and (\ref{main-eq-2}), we can take saturation and
one of the pressures as independent variables; for example $S_g$ and
$p_l$, and  for the other terms the following functional
dependencies:
\[ B_l(p_l), B_g(p_g), R_s(p_l), R_v(p_g), \mu_l(p_l), \mu_g(p_g), kr_w(S_g),
kr_g(S_g).
\]
According to their definition, $F$, $\rho_{g}^{std}$,
$\rho_{l}^{std}$ and $D^i_\alpha$ ($\alpha\in \{l,g\}, i\in \{w,h\}$) are
constants, and $\Phi$ and $\Kb$ are depending only on space
position.

 \subsubsection{Unsaturated flow}
\label{sec:unsaturated}

Equations  (\ref{main-eq-1}) and (\ref{main-eq-2}), with  saturation and
 pressure as unknowns, are valid if there is no missing phase (saturated flow); but if
one of the phases is missing (unsaturated flow), equations and
unknowns have to be adapted. There are two possible {\em
unsaturated} cases, according to either the gas phase or the liquid
phase disappears :
\begin{enumerate}
\item
\underline{Gas phase missing} (Hydrogen totally dissolved in
water): then we have $S_g=0$, $S_l=1$. Generalized gas phase Darcy's
velocity is equal to zero since $kr_g(0)=0$. Independent variables
are now $p_l$, the liquid phase pressure, and $R_s$, the solution
gas/liquid phase Ratio; then we must write $B_l=B_l(p_l,R_s)$ and
$\mu_l=\mu_l(p_l, R_s)$, for $0\leq R_s\leq \hat{R}_s(p_l)$, where
$\hat{R}_s(p_l)$ is equilibrium solution gas/liquid phase ratio.
\item
\underline{ Liquid phase missing}: then, $S_l=0$, $S_g=1$.
Generalized liquid phase Darcy's velocity is zero since
$kr_w(S_g=1)=0$; independent variables are now $p_g$, the gas phase
pressure, and  $R_v$ the  water vapor/gas phase ratio, become the
new independent variables. However pressure $p_l$ can be kept as
independent variable since $p_g$ could be expressed through the
capillary pressure law (\ref{capillary}). We must also write
$B_g=B_g(p_g,R_v)$ and $\mu_g=\mu_g(p_g, R_v)$, for $0\leq R_v\leq
\hat{R}_v(p_g)$, where $\hat{R}_v(p_g)$ is the equilibrium water
vapor/gas phase ratio.
\end{enumerate}
These above conditions can be summarized as
\begin{align}
  S_g \geq 0,\quad \hat{R}_s(p_l) - R_s \geq 0,\quad (\hat{R}_s(p_l) - R_s )S_g=0,\\
  S_l \geq 0,\quad \hat{R}_v(p_g) - R_v \geq 0,\quad (\hat{R}_v(p_g) - R_v ) S_l =0.
\end{align}
\begin{remark}
In this last section, Phase Equilibrium Black-oil model,
we did not take in account a possible interplay between dissolution
and capillary pressure. $\;\Box$
\end{remark}

\subsection{Thermodynamical equilibrium Henry-Raoult model}

Another way of closing the system of equations (\ref{saturations}),
(\ref{capillary}), (\ref{eq-1}) and (\ref{eq-2}) is  to use phase
thermodynamical properties for characterizing equilibrium . We use
first ideal gas law and Dalton law,
\begin{equation}
 p_g = p_g^w + p_g^h,
\label{Dalton}
\end{equation}
where $p_g^w$ and $p_g^h$ are the vaporized water and hydrogen
partial pressures in the gas phase; and
\begin{equation}
 p_g^w = \frac{\rho_g^w}{M^w} RT,\quad  p_g^h = \frac{\rho_g^h}{M^h} RT .
\label{ideal}
\end{equation}
$T$ is the temperature, $R$  is universal gas constant
and $M^w$, $M^h$ are the water and hydrogen molar masses.

Next, we apply 
{\em Henry's} and {\em Raoult's laws} which
say that,
 at equilibrium, the vapor pressure of a substance  varies linearly with its mole fraction in solution.
 In {\em Henry's law} the constant of proportionality  
 is obtained 
 by experiment  and in {\em Raoult's law} the constant 
 is the  pressure of
 the component 
 in its pure state.
  Here we will assume,
  for simplicity, that  the quantity of dissolved hydrogen
 in the liquid is small; then these laws reduce to the linear {\em Henry's
law}, which says that the amount of gas dissolved in a given volume
of the liquid phase is directly proportional to the partial pressure
of that same gas in the gas phase:
\begin{equation}
 \rho_l^h= H(T) M^h p_g^h ,   \label{Henry}
\end{equation}
where $H(T)$ is the Henry's law constant, depending only on the
temperature. For the liquid  phase,
we apply  {\em Raoult's law} which says that the \textit{water
vapor pressure} is equal to the vapor pressure of the pure solvent,
at given temperature, multiplied by the mole fraction of the
solvent. The water vapor partial pressure of the pure solvent
depends only on the temperature and therefore is a constant, denoted
here by $\hat{p}_g^w(T)$, so we have from definition
(\ref{molar-concentrations})
\begin{equation}
   p_g^w = \hat{p}_g^w(T) X_l^w = \hat{p}_g^w(T)
   \frac{\rho_l^w}{\rho_l^w +(M^w/M^h)\rho_l^h}.
\label{Raoult}
\end{equation}
Further on, we can  include in  formula (\ref{Raoult}) the presence
of capillary pressure, by using {\em Kelvin}'s equation (see
\cite{Abbas}) which gives:
\begin{equation}
   p_g^w =  \hat{p}_g^w(T)
   \frac{\rho_l^w}{\rho_l^w +(M^w/M^h)\rho_l^h} e^{-M^w p_c/(RT\rho_l)}.
\label{Raoult-Kelvin}
\end{equation}
If now, to equations (\ref{Dalton})-(\ref{Henry}) and
(\ref{Raoult-Kelvin})  we add the relations
\begin{equation}
\rho_l^h + \rho_l^w =\rho_l,\quad \rho_g^h + \rho_g^w =\rho_g,
\label{summ}
\end{equation}
and the  water compressibility, defined by
\begin{equation}
    \rho_l^w = \frac{\rho_{l}^{std}}{B_l(p_l)};
\label{water-compress}
\end{equation}
then, we have 8 equations: (\ref{Dalton}), (\ref{ideal})$_{1,2}$,
(\ref{Henry}),  (\ref{Raoult-Kelvin}), (\ref{summ})$_{1,2}$ and
(\ref{water-compress}) and 10 unknowns:
\[ p_l, p_g,  p_g^w, p_g^h, \rho_l^h, \rho_l^w, \rho_l,  \rho_g^h, \rho_g^w, \rho_g.\]

 We may, for instance, parametrize all these 10 unknowns,
 by the two phase pressures $p_l$ and $p_g$. For this we should
combine the Henry law (\ref{Henry}), the Raoult-Kelvin law
(\ref{Raoult-Kelvin}) and (\ref{Dalton}) leading to the system of
two equations for $p_g^h$ and $p_g^w$:
\begin{align*}
    p_g^w &= \hat{p}_g^w(T)
   \frac{\rho_l^w}{\rho_l^w +(M^w H(T))p_g^h}  e^{-M^w (p_g -p_l)/(RT(\rho_l^w+ M^h H(T) p_g^h)}\\
    p_g &= p_g^w + p_g^h.
\end{align*}
It is easy to show that this above system of two equations  has a
unique  solution $ p_g^w, p_g^h >0$
for any $p_g \geq \hat{p}_g^w(T)$; and solving this system we obtain
\[
   p_g^w = f(p_g, p_l),\quad p_g^h = g(p_g, p_l).
\]
By (\ref{ideal})$_{1,2}$ we can then  write $ \rho_g^h$,
$\rho_g^w$ and $\rho_g$ as functions of $p_l$ and $p_g$, and by
(\ref{Henry}) and (\ref{water-compress}) we can finally express
$\rho_l^h$, $\rho_l^w$ and $\rho_l$ as functions of the phase pressures.
\begin{remark}
The gas phase will appear only if there is a sufficient quantity of
dissolved hydrogen in the liquid phase, and this quantity is exactly
the dissolved gas quantity, at equilibrium, given by  Henry's law.
But
when the dissolved gas (hydrogen) quantity is smaller than the
quantity of hydrogen at equilibrium, then the Henry
law does not apply; and $S_g$ is then equal to zero and could not be
taken as unknown. In this situation, instead of saturation, we may
take $\rho_l^h$  as independent variable.
We notice that Henry-Raoult model based on thermodynamical equilibrium leads
to a similar concepts as the ones developed for establishing Black Oil
model for reservoir modeling.
 $\;\Box$
\end{remark}
\begin{remark}
In the Henry-Raoult model, if there is no vaporized water,
$p_g=p_h^g$, the criteria for non saturated flow is simple and reads
 \begin{equation}
 \frac{\rho_l^h}{M^h H(T)} <p_g= p_l +p_c(0),
 \label{seuil}
 \end{equation}
  if there is a threshold pressure, i.e. $p_c(0)\neq 0$.
  The same criterion can be used also if there is vaporized water,
  as long as the pressure in porous medium  is much larger than  vaporized
  water pressure. \label{rem_nonequ}
\end{remark}
\begin{remark}
  In Henry-Raoult model, above, we were assuming for simplicity,  no
  complete evaporation of the water.
\end{remark}

\subsubsection{Comparison of Equilibrium models}

We will now consider in more details a special case, where we may,
from the Henry-Raoult model,
get back the Black oil model. In  both models we will then use diffusive fluxes
developed in Section~\ref{black-oil}.
For simplicity, we will neglect in (\ref{Raoult}) influence of capillary pressure
and  solved gas on water vapor partial pressure,
leading to
$ p_g^w = \hat{p}_g^w(T)$ being a constant.
 Then we have from (\ref{Henry}), (\ref{water-compress}) and (\ref{summ})
\begin{align}
    \rho_l =   \rho_l^h + \rho_l^w = H(T)M^h  p_g^h + \frac{\rho_l^{std}}{B_l(p_l)}
    = \frac{\rho_l^{std} + B_l(p_l) H(T)M^h  (p_g-\hat{p}_g^w(T))}{B_l(p_l)} ,
    \label{Rs-approx}
\end{align}
and also from (\ref{ideal}),
\begin{align}
    \rho_g & =  \rho_g^h+ \rho_g^w    =  \frac{M^h}{RT} (p_g -\hat{p}_g^w(T))
+  \frac{M^w}{RT} \hat{p}_g^w(T).
 \label{Rs-approx-1}
\end{align}
Therefore, equations (\ref{Rs-approx}) and (\ref{Rs-approx-1}) can
be written in a form similar to (\ref{density-petrol})-(\ref{density-petrol-gas})
by defining the gas  formation volume factor $B_g$, solution gas/liquid phase ratio
$R_s$ and vapor water/gas phase ratio $R_v$, from above thermodynamical relations:
\begin{align}
  \quad R_s =   B_l(p_l)\frac{ H(T)M^h}{\rho_g^{std}} ( p_g -\hat{p}_g^w(T) )  \label{phase-2-1}\\
  {B_g}  =   \frac{RT\rho_g^{std}}{M^h( p_g -\hat{p}_g^w(T))}, \quad R_v =
    \frac{1}{F} \frac{\hat{p}_g^w(T)}{p_g -\hat{p}_g^w(T)}, \label{phase-2-2}
\end{align}
where $F$ is given by (\ref{F}).

 Compared to the Black-Oil model,
Section~\ref{black-oil}, here, it is clear from (\ref{phase-2-1})
that  solution gas/liquid phase ratio $R_s$  depends on $p_l$ and
$p_g$ and not only on $p_l$.

\begin{remark}
For the case without any water  vapor, the corresponding thermodynamical model
is obtained simply by taking $\hat{p}_g^w(T)=0$ in (\ref{phase-2-1}), (\ref{phase-2-2})
and $R_v=0$ in equations (\ref{main-eq-1}) and (\ref{main-eq-2}). In the same way, if
 we neglect dissolved hydrogen, the corresponding thermodynamical model is
 obtained simply by taking $H(T)=0$
 in (\ref{phase-2-1}), (\ref{phase-2-2}), and $R_s=0$ in
 (\ref{main-eq-1}) and (\ref{main-eq-2}).
  $\;\Box$
\end{remark}

\subsubsection{Model assuming water incompressibility and no water vaporization}

In the Henry-Raoult model we assume now that the water is
incompressible and  the water vapor quantity is neglectful; i.e. the
gas phase contains only hydrogen, then (see Remark~\ref{rem_nonequ})
   $p^{h}_g\equiv p_g$ and $p_g^w=\hat{p}_g^w(T)=0$ in (\ref{phase-2-2}) leading
to $R_v \equiv 0$.
Formulas (\ref{phase-2-1}), (\ref{phase-2-2}), could be rewritten as
\begin{align}
 B_l \equiv 1,&\quad R_s  =   C_h p_g,\quad
    \frac{1}{B_g}  =   C_v p_g; \label{phase-2-3}
\end{align}
where we have denoted
\begin{align}
    C_h= \frac{H(T)M^h}{\rho_g^{std}},\quad C_v = \frac{M^h}{RT \rho_g^{std}}; \label{phase-2-4}
\end{align}
and equations (\ref{density-petrol})-(\ref{density-petrol-gas}) become:
\begin{align}
  \rho_l(R_s) = \rho_{l}^{std} + R_s\rho_{g}^{std}, \quad
  \rho_g(p_g) =  C_v\rho_{g}^{std} p_g .  \label{phase-2-4-1}
\end{align}
Similarly to constant $F$ defined by (\ref{F}), we use the density
ratio
\begin{equation}
  G = \frac{\rho_{l}^{std}}{\rho_{g}^{std}},
\label{densityratio}
\end{equation}
and equations (\ref{main-eq-1}),  (\ref{main-eq-2}) reduce to
\begin{align}
  \Phi\dert{S_l} +
  \dv \left( \ql  -\frac{1}{G}\Jb \right)={\cal F}^w/\rho_{l}^{std},
  \label{consSl}\\[2mm]
  \Phi\dert{}( S_l R_s   + C_v p_g S_g  )
  +\dv \Big(R_s \ql +  C_v p_g \qg  + \Jb\Big)
  = {\cal F}^h/\rho_{g}^{std} , \label{h2-mass-cons}\\
  \ql = -\Kb\frac{kr_l}{\mu_l}
  \left( \gr p_l - (\rho_{l}^{std} + R_s\rho_{g}^{std}) \gb\right),
  \quad
  \qg = - \Kb\frac{kr_g}{\mu_g} \left( \gr p_g -C_v\rho_{g}^{std} p_g \gb\right),
  \label{darcy-vel}\\
  \Jb = -  \frac{\Phi S_l F}{R_s + F} D_l^h \gr R_s,  \label{eq:5}
\end{align}
where we have denoted $ \boldsymbol{\phi}_l^h = \Jb $
and used Remark~\ref{rem3} to get $F D_l^h = G D_l^w$, from which it follows
$\boldsymbol{\phi}_l^w = -\Jb/G.$

Here there is only hydrogen in phase gas, $p^h_g=p_g$ and Henry's
law assumes thermodynamical equilibrium in which  quantity of
dissolved hydrogen is proportional to  gas phase pressure. In
saturated case (where two phases are present)  Henry's law reads
$R_s = C_h p_g$,
and we can then work with variables $p_g$ and $S_l$ in equations
(\ref{consSl})--(\ref{eq:5}).

When
the gas phase disappears,
the gas pressure drops
to the liquid pressure augmented by entry pressure, $p_g=p_l+p_c(0)$,
 the liquid can contain any
quantity of dissolved hydrogen $\rho^{h}_g$ between zero and
$C_h \rho_{g}^{std} (p_l+p_c(0))= H(T)M^h (p_l+p_c(0))$, from
Henry's law (see (\ref{Henry}) and definition (\ref{phase-2-4})).

And then, when the gas phase is absent (one of the unsaturated
cases) $S_g=0$, we will replace saturation, as we did in the
Black-Oil model for the same unsaturated case in
Section~\ref{sec:unsaturated}, by a new variable $R_s$ (see
definitions (\ref{Henry}), (\ref{phase-2-1}), (\ref{phase-2-4})), such that
 $$ R_s \rho_{g}^{std}= \rho^h_l $$
 is the mass density of dissolved hydrogen in the liquid phase.

Assuming at standard conditions the gas phase contains only hydrogen
(hydrogen component mass$\approx$gas phase mass) and the liquid
phase contains only water, with  water incompressibility and no
vaporized water, we see that, in definition (\ref{phase-2-1}), $R_s=
\rho_{l}^{h}/\rho_{g}^{std}\simeq \Delta V_g^{std}/\Delta V_l^{std}
$ which is exactly the definitiion given in the Black-Oil model in
Section~\ref{black-oil}.

 The physical meaning of $R_s$  then stays the same
either in Henry-Raoult or in Black-Oil models even in the
unsaturated case. Moreover, from mass conservation law it follows
that the dissolved hydrogen mass density $R_s\rho_{g}^{std} $ is
continuous when the gas phase vanishes and this can be expressed as
an unilateral condition:
\begin{align*}
    0\leq S_g \leq 1,\quad 0\leq R_s \leq C_h p_g,\quad S_g (C_h p_g - R_s ) = 0.
\end{align*}

In {\em unsaturated region}, where $S_l=1$ (that is $S_g =0$) we
replace variable $S_l$ by $R_s$, and equations
(\ref{consSl})--(\ref{eq:5}) degenerate to:
 \begin{align}
  & \dv \left(   \ql -\frac{1}{G}\Jb \right)={\cal F}^w/\rho_{l}^{std}; \label{un:eq:1}\\
  \Phi\dert{R_s} & + \dv \Big(
  R_s \ql  +\Jb \Big) = {\cal F}^h/\rho_{g}^{std} ;\label{un:eq:2}\\[2mm]
  \ql &= -   \Kb\lambda_l(1)
            \left( \gr p_l - (\rho_{l}^{std} + R_s \rho_{g}^{std}) \gb\right); \\[2mm]
  \Jb &= -  \frac{\Phi F }{R_s + F} D_l^h \gr R_s.
   \label{un:eq:4}
   \end{align}

\subsection{Saturated/unsaturated state, general formulation }

Finally in the saturated region we used $p_l$ and $S_g$ as variables
in (\ref{consSl})--(\ref{eq:5})  but in  unsaturated region we
should use other variables, $p_l$ and $R_s$, in
(\ref{un:eq:1})--(\ref{un:eq:4}). In order to avoid the change of
variables and equations in different regions as above we prefer to
introduce a new variable \begin {equation} X=  (1-S_g) R_s   + C_v
p_g S_g; \label{newvar}
 \end{equation}
 in view to make equation (\ref{h2-mass-cons})
parabolic in $X$.

 This new variable $X$ is well defined both in saturated and unsaturated
regions. Moreover, from (\ref{phase-2-3}) and (\ref{Henry}),
    $R_s={\rho_{l}^{h}}/{\rho_{g}^{std}}$; from
(\ref{phase-2-4-1}) $C_v p_g={\rho_{g}}/{\rho_{g}^{std}}$ and since
$\rho_{g}=\rho_{g}^{h}$, this new variable $X$
 is a "\textbf{normalized total hydrogen mass density}":
$X =(S_l \rho_{l}^{h}+ S_g \rho_{g}^{h})/{\rho_{g}^{std}}$.
It is easy then  to see that parabolicity is possible
only if we take liquid pressure $p_l$ as other independent variable.
Knowing that in saturated case, $S_g>0$, $R_s=C_h p_g$ from Henry's
law, we may write $X$ defined in (\ref{newvar})  as:
\begin{align}
    X = \begin{cases}
        (C_h (1-S_g)   + C_v S_g )   (p_l + p_c(S_g) ) & \text{if } S_g >0\\
        R_s & \text{if } S _g = 0.
    \end{cases} \label{def-X}
\end{align}
Since for capillary pressure  defined  as a function of gas saturation
we have $p_c'(S_g)
> 0$, and since usually, like for hydrogen, in (\ref{phase-2-4})
    $\omega = C_v/C_h > 1$
we get the following bounds:
\begin{align}
  a(S_g) = C_h (1-S_g)   + C_v S_g  \in [ C_h, C_v],\quad a'(S_g) = C_v - C_h =C_{\Delta} > 0.
  \label{a:pos}
\end{align}
Then for any  $S_g > 0$ from (\ref{def-X}):
\begin{align*}
    \der{X}{S_g} = C_{\Delta}  (p_l + p_c(S_g) ) +   a(S_g) p_c'(S_g)  > 0,
\end{align*}
and therefore for each $p_l >0$ we can find inverse function $S_g=S_g(p_l,X)$
which satisfies
\begin{align*}
    \der{S_g}{X} > 0\quad\text{for } S_g > 0 .
\end{align*}
After taking derivatives with respect to $p_l$ and $X$ of (\ref{def-X})
we obtain:
\begin{align}
    \der{S_g}{p_l} =  -\frac{a(S_g)^2  \chi(p_l,X)}{C_{\Delta} X +a(S_g)^2 p_c'(S_g)},\quad
    \der{S_g}{X} =  \frac{a(S_g)  \chi(p_l,X)}{C_{\Delta} X +a(S_g)^2 p_c'(S_g)},\label{pos:def:S}
\end{align}
where $\chi(p_l,X)$ is characteristic function of the set $\{ X > C_h (p_l +p_c(0)) \}$.
In (\ref{pos:def:S}), we remark from (\ref{a:pos}) and definition (\ref{capillary}) that
$\partial S_g/\partial p_l \leq 0$.

Let us note that property
\begin{align}
    p_c'(S_g=0) = +\infty, \label{pc:infty}
\end{align}
of van Genuchten $p_c$ functions, leads to continuity of the two above partial
derivatives since we have
\begin{align*}
    \lim_{S_g\to 0}  \der{S_g}{p_l} =  \lim_{S_g\to 0}   \der{S_g}{X} =0.
\end{align*}
We now  introduce auxiliary function  $N(p_l,X)$ defined as
\begin{align*}
    N(p_l,X) =  \frac{C_{\Delta} X}{C_{\Delta} X +a(S_g)^2 p_c'(S_g)}\chi(p_l,X)
    \in [0,1),
\end{align*}
which verifies
\begin{align}
\chi(p_l,X) +  p_c'(S_g)\der{S_g}{p_l}  =  N(p_l,X),\quad
 p_c'(S_g)\der{S_g}{X}  =  \frac{1-N(p_l,X)}{a(S_g)} \chi(p_l,X). \label{N-fun:aux}
\end{align}
Note that function $N(p_l,X)$ is continuous under condition (\ref{pc:infty}).

Darcy's fluxes in (\ref{darcy-vel}) and diffusive flux (\ref{eq:5}) now take the form,
with $p_g = p_l + p_c(S)$,
\begin{align}
  \ql &= -   \Kb\lambda_l(S_g)
            \left( \gr p_l  - (\rho_{l}^{std} + R_s(p_l,X)\rho_{g}^{std}) \gb\right)\\
  \qg &= -  \Kb\lambda_g(S_g),
           \left( \gr p_l+\gr p_c(S_g) -C_v \rho_{g}^{std} p_g(p_l,X)  \gb\right),
           \label{new_flux}\\
  \Jb &= -  \frac{\Phi (1-S_g) F}{R_s(p_l,X) + F} D_l^h \gr R_s(p_l,X); \label{new:eq:5}
\end{align}
where $S_g$ is a function of $p_l$ and $X$ and where $R_s$, defined
as in (\ref{phase-2-3}), can now be expressed in both saturated and
unsaturated region as a function of new variables $p_l$ and $X$:
\begin{align*}
 R_s(p_l,X)  = \min(C_h  p_g(p_l,X),X),\quad
 p_g(p_l,X)  = p_l+p_c(S_g(p_l,X)).
\end{align*}
After expanding the capillary pressure gradient in
(\ref{new_flux}) and the gradient of $R_s$ in (\ref{new:eq:5}), we may write:
\begin{align*}
 \qg =& -  \Kb\lambda_g(S_g) \left( [1+ p_c'(S_g) \der{S_g}{p_l}]\gr p_l
     + p_c'(S_g) \der{S_g}{X}\gr X -C_v \rho_{g}^{std}  p_g(p_l,X)  \gb\right),\\
  \Jb =& -  \frac{\Phi (1-S_g(p_l,X)) F}{R_s(p_l,X) + F}
    D_l^h C_h \chi( 1 +  p_c'(S_g)\der{S_g}{p_l} )\gr p_l \\
& -  \frac{\Phi (1-S_g(p_l,X)) F}{R_s(p_l,X) + F}
      D_l^h C_h  p_c'(S_g)\der{S_g}{X} \gr X - \frac{\Phi  F}{ X + F} D_l^h  (1-\chi)\gr X.
\end{align*}
From (\ref{comp:flux}) water component flux $\boldsymbol{\phi}^{w}$ and
hydrogen component flux $\boldsymbol{\phi}^{h}$, with the assumptions of incompressible water
and absence of water vapor, reduce to
\begin{align}
\boldsymbol{\phi}^{w} &= \ql +  \boldsymbol{\phi}_l^w =
-(\tilde A_{1,1}\gr p_l +\tilde A_{1,2}\gr X +\tilde B_1), \label{w-tot}\\
\boldsymbol{\phi}^{h} &= R_s \ql +C_v p_g \qg +\boldsymbol{\phi}_l^h =
-( A_{2,1}\gr p_l +A_{2,2}\gr X +B_2),\label{h-tot}
\end{align}
where the coefficients $A_{ij}(p_l,X)$ and $B_i(p_l,X)$ are given by the following
formulas
(note that from (\ref{eq:5}) $\boldsymbol{\phi}_l^h =\Jb$ and $\boldsymbol{\phi}_l^w=-\Jb/G$):
\begin{align}
    \tilde A_{11}(p_l,X) =&    \Kb\lambda_l(S_g) -
     \frac{\Phi (1-S_g) F}{(R_s + F)G} D_l^h C_h N  ,  \label{coef:in:1}\\
    \tilde A_{1,2}(p_l,X) =&
    -  \frac{\Phi (1-S_g) F }{(R_s + F)G}\frac{1-N}{a(S_g)} D_l^h C_h, \\
    A_{2,1}(p_l,X) =&
    \Kb\lambda_l(S_g)R_s +\Kb\lambda_g(S_g) C_v p_g N
        +\frac{\Phi  (1-S_g)F}{R_s + F} D_l^h C_h N,   \\
    A_{2,2}(p_l,X) =&    \Kb\lambda_g(S_g) \frac{1-N}{a(S_g)}C_v p_g
                   +\frac{\Phi (1-S_g) F}{R_s + F}  \frac{1-N}{a(S_g)} D_l^h C_h, \\
    \tilde B_1(p_l,X) = &
    -   \Kb\lambda_l(S_g)  [\rho_{l}^{std} + R_s \rho_{g}^{std}] \gb,\\
    B_2(p_l,X) = & -\Kb\lambda_l(S_g)  R_s [\rho_{l}^{std} + R_s \rho_{g}^{std}]\gb
                 - \Kb\lambda_g(S_g)C_v^2 \rho_{g}^{std} p_g^2 \gb;  \label{coef:in:2}
\end{align}
where we have used (\ref{N-fun:aux}) and $\lambda_g(S_g) (1-\chi)=0$.

Equations (\ref{consSl})--(\ref{eq:5}) become:
\begin{align}
 -\Phi\der{S_g}{p_l}\dert{p_l} &
 - \dv \left( \tilde A_{1,1}\gr p_l +\tilde A_{1,2}\gr X +\tilde B_1\right)
  -\Phi\der{S_g}{X}\dert{X}    ={\cal F}^w/\rho_{l}^{std}
  \label{eq:1-3}\\
  \Phi\dert{X} & - \dv \Big( A_{2,1}\gr p_l +A_{2,2}\gr X +B_2\Big)
      = {\cal F}^h/\rho_{g}^{std}.\label{eq:2-3}
\end{align}

The gain in this form is that (\ref{eq:2-3}) is a parabolic equation for $X$ since
\begin{align*}
    A_{2,2}(p_l,X)\boldsymbol{\xi}\cdot \boldsymbol{\xi} =&
    \Kb\boldsymbol{\xi}\cdot \boldsymbol{\xi}\lambda_g(S_g) \frac{1-N}{a(S_g)} C_v p_g
                   + \frac{\Phi (1-S_g) F}{R_s + F}\frac{1-N}{a(S_g)} D_l^h C_h
                   |\boldsymbol{\xi}|^2
\end{align*}
is strictly positive in the whole domain if the diffusion and capillary pressure are
not neglected.

If we eliminate diffusive terms from  equation (\ref{eq:1-3}) (the
"pressure equation")  by forming an equation for
total flow $\boldsymbol{\phi}_{tot}$, defined in (\ref{tot-tot}), that is summing equation
(\ref{eq:2-3}) and equation (\ref{eq:1-3}), we obtain
\begin{align}
\boldsymbol{\phi}_{tot} = G \boldsymbol{\phi}^{w} +
\boldsymbol{\phi}^{h} = (G+R_s) \ql + C_v p_g \qg = -(  A_{1,1}\gr
p_l + A_{1,2}\gr X +  B_1),\label{tot-tot}
\end{align}
where $G$ is the density ratio defined in (\ref{densityratio}), and coefficients
$A_{1,1}, A_{1,2}$ and $B_1$ are given by:
\begin{align}
    A_{1,1}(p_l,X) =& G  \tilde  A_{1,1} +   A_{2,1}
                       = \Kb\lambda_l(S_g)(G+R_s) +\Kb\lambda_g(S_g) C_v p_g N,
    \label{coef:in:3} \\
    A_{1,2}(p_l,X) =& G \tilde  A_{1,2} + A_{2,2}
                       =    \Kb\lambda_g(S_g) \frac{1-N}{a(S_g)} C_v p_g ,\\
    B_1(p_l,X) = & G \tilde B_1 + B_2 =
    -\Kb\lambda_l(S_g) (G+ R_s) [\rho_{l}^{std} + R_s \rho_{g}^{std}]\gb
                 - \Kb\lambda_g(S_g)C_v^2 \rho_{g}^{std} p_g^2 \gb.
 \label{coef:in:4}
\end{align}
Now the "pressure equation" (\ref{eq:1-3}) is transformed to:
\begin{align}
-G\Phi\der{S_g}{p_l}\der{p_l}{t} - \dv ( A_{1,1}\gr p_l + A_{1,2}\gr X &+  B_1 )
    + \Phi(1-G\der{S_g}{X})\der{X}{t}  \label{eq:1-3-tot} \\
     &= G {\cal F}^w/\rho_{l}^{std} +{\cal F}^h/\rho_{g}^{std}.\nonumber
 \end{align}
With this last formulation,  we see that the "pressure equation"
(\ref{eq:1-3-tot}) is parabolic/elliptic equation in $p_l$ since
\begin{align*}
     A_{11}(p_l,X)\boldsymbol{\xi}\cdot \boldsymbol{\xi}  =
     \Kb\boldsymbol{\xi}\cdot \boldsymbol{\xi} \lambda_l(S_g)(G+R_s)  +
      \Kb\boldsymbol{\xi}\cdot \boldsymbol{\xi}\lambda_g(S_g) C_v p_g N,
\end{align*}
is strictly positive, independently of presence of diffusion
or capillary forces, and the coefficient in front of $\partial p_l/\partial t$
is  positive since, as we remarked at (\ref{pos:def:S}),
$\partial S_g/\partial p_l \leq 0$.

Finally, the transport of water and hydrogen is described by differential equations
(\ref{eq:1-3-tot}) and (\ref{eq:2-3}) which are rewritten here, using
(\ref{h-tot}) and (\ref{tot-tot}), in the form
\begin{align}
\Phi\der{}{t}(X-G S_g(p_l,X))&  + \dv \left( \boldsymbol{\phi}_{tot}\right)
     = G {\cal F}^w/\rho_{l}^{std} +{\cal F}^h/\rho_{g}^{std},\label{eq:1-3-fin}\\
  \Phi\dert{X} & + \dv \Big(\boldsymbol{\phi}^h \Big)
      = {\cal F}^h/\rho_{g}^{std},\label{eq:2-3-fin}
\end{align}
where the fluxes are given by (\ref{h-tot}) and (\ref{tot-tot}),
while the coefficients are given by formulas   (\ref{coef:in:1})--(\ref{coef:in:2})
and (\ref{coef:in:3})--(\ref{coef:in:4}).

\subsubsection{Boundary conditions}

Equations (\ref{eq:1-3-fin}) and (\ref{eq:2-3-fin}), given in porous
domain  $\Omega$,   must be complemented by initial and boundary
conditions. Following  (\cite{ChJa}), we assume that the boundary
$\partial\Omega$ is divided in several disjoint parts: impervious,
inflow and outflow boundaries. We present now a set
of standard boundary conditions on each of these  boundary parts.

\begin{itemize}
\item  On  impervious boundary $\Gamma_{imp}$ we take  Neumann conditions:
\begin{equation}
\boldsymbol{\phi}_{tot}\cdot \boldsymbol{\nu} = 0
 \;\;\text{and }\;\;
   \boldsymbol{\phi}^{h}\cdot \boldsymbol{\nu}= 0 .
\end{equation}
\item On inflow boundary when pure water is injected, we impose for hydrogen
component $X=0$ and for liquid pressure either $p_l=p_{l,in}$
or $\boldsymbol{\phi}_{tot}\cdot \boldsymbol{\nu} = Q_d .$
\item On inflow boundary,  when  pure gas is injected we have
$\boldsymbol{\phi}^{w}\cdot \boldsymbol{\nu}= 0$ and we can impose
 for  the pressure, total injection rate
 which is then equal to gas injection rate:
\begin{equation}
    \boldsymbol{\phi}_{tot}\cdot \boldsymbol{\nu}= Q_{d}^h
    =\boldsymbol{\phi}^{h}\cdot \boldsymbol{\nu}.  
\end{equation}
\item On the outflow boundary, when liquid is displaced by gas we have possibility to
impose for gas phase $X=0$ and for
liquid pressure $p_l =p_{l,out}$, only before gas reaches this
outflow boundary (breakthrough time).
 Alternatively, with the same Dirichlet condition for liquid pressure,
 for gas saturation we can set either Neumann condition
 $ \nabla S_g\cdot \boldsymbol{\nu}=0$,
 or Dirichlet condition
 $S_g=0$.
\end{itemize}

\subsection{Model without capillary pressure, diffusion or gravity}

When capillary pressure is neglected,  we have only one pressure
$p_g=p_l=p$, then definition (\ref{def-X}) of variable $X$
simplifies to:
\begin{align}
    X = \begin{cases}
        (C_h (1-S_g)   + C_v S_g )  p  & \text{if } S_g >0\\
        R_s & \text{if } S _g = 0,
    \end{cases} \label{newvar1}
\end{align}
and partial derivatives (\ref{pos:def:S}) can be calculated
explicitely ,
\begin{align*}
    \der{S_g}{p} =  -\frac{X}{C_{\Delta} p^2 } \chi(p,X),\quad
    \der{S_g}{X} =  \frac{1}{C_{\Delta} p} \chi(p,X),
\end{align*}
where $\chi(p,X)$ is the characteristic function of saturated region, as before.

When we neglect capillary pressure, and also diffusive and gravity
fluxes, system (\ref{eq:1-3-fin}),  (\ref{eq:2-3-fin}) reduces to
 the following two equations
\begin{align}
\Phi\der{}{t}(X-G S_g(p,X))&  + \dv \left( \boldsymbol{\phi}_{tot} \right)
     = G {\cal F}^w/\rho_{l}^{std} +{\cal F}^h/\rho_{g}^{std}\label{eq:1-3x}\\
  \Phi\dert{X} & + \dv \Big( f^h(p,X) \boldsymbol{\phi}_{tot} \Big)
      = {\cal F}^h/\rho_{g}^{std},\label{eq:2-3s}
     \end{align}
  where
\begin{align}
       \boldsymbol{\phi}_{tot}= - \Lambda_{tot}(p,X)\Kb\gr p, \quad
       \boldsymbol{\phi}^{h} =f^h(p,X) \boldsymbol{\phi}_{tot}.\label{eqfluxH}
\end{align}
Total mobility $ \Lambda_{tot}(p,X)$ and hydrogen fractional flow
functons $f^h(p,X)$ are defined by:
\begin{align*}
  \Lambda_{tot}(p,X)&=\lambda_l(S_g)(G+R_s(p,X)) +\lambda_g(S_g) C_v p , \\
    \Lambda^h(p,X)&=\lambda_l(S_g)R_s(p,X) +\lambda_g(S_g) C_v p, \quad
  f^h(p,X) =\frac{\Lambda^h(p,X)}{\Lambda_{tot}(p,X)}  .
\end{align*}
System (\ref{eq:1-3x})--(\ref{eqfluxH}) is very close to  immiscible
two-phase system (see for instance \cite{ChJa}); the only difference
in hydrogen transport equation (\ref{eq:2-3s}) is the presence of
$R_s(p,X)$ in fractional flow function $ f^h(p,X)$, and if we write
 pressure equation  (\ref{eq:1-3x}) in the form
\begin{align}
G\chi\Phi\frac{X}{C_{\Delta} p^2} \der{p}{t}&  + \dv \left( \boldsymbol{\phi}_{tot} \right)
    + \Phi (1 - \frac{ G \chi}{C_{\Delta} p} )\dert{X}
    ={G {\cal F}^w/\rho_{l}^{std} +\cal F}^h/\rho_{g}^{std},
\end{align}
then we see that the presence of dissolved gas introduces additional
"source" term in total flow equation (\ref{eq:1-3x}), namely $\Phi
(1 -{ G \chi}/(C_{\Delta} p) )\partial{X}/\partial t$.
\begin{remark}
Considering the boundary conditions defined in the previous section;
on the boundary part where there is hydrogen outflow, to impose a
Dirichlet condition on $X$ will lead to a boundary layer.
\end{remark}

\section{Numerical simulations}

This section presents two test cases and their simulations using the new formulation given by system (\ref{eq:1-3-fin}), (\ref{eq:2-3-fin}).
 Both test cases are not build to correspond with particular real situations but rather to illustrate the gas appearance phenomenon.
  Assuming horizontal two dimensional problems gravity effects are neglected in both cases. The first test case is a one dimensional
  like situation where hydrogen is injected through an inflow boundary and the second test case is a two dimensional situation where hydrogen
  is injected via a volume source term.

\subsection{Setting test cases}

\subsubsection{Physical data}

In the two test cases we consider the same isotropic porous medium with a uniform absolute permeability 
tensor $\Kb=k$ where $k$ is scalar and a uniform porosity $\Phi$.
The capillary pressure function, $p_c$, is given by the van Genuchten model (see \cite{VG80}) and 
relative permeability functions, $kr_l$ and $kr_g$, are given by the van Genuchten-Mualem model 
(see \cite{VG80} and \cite{Mualem}). According to these models we have~:
\begin{eqnarray*}
    & p_c = P_r\left(S_{le}^{-1/m}-1\right)^{1/n} ,\
    kr_l = \sqrt{S_{le}}\left( 1-(1-S_{le}^{1/m}) \right)^2 
    \mbox{and}\
    kr_g=\sqrt{1-S_{le}}\left( 1-S_{le}^{1/m} \right)^{2m}&\\
    \lefteqn{\mbox{with}\quad
    S_{le}=\frac{S_l-S_{lr}}{1-S_{lr}-S_{gr}}
    \quad\mbox{and}\quad m=1-\frac{1}{n}}
\end{eqnarray*}
where parameters $P_r$, $n$, $S_{lr}$ and $S_{gr}$ depend on the porous medium.
Values of parameters describing the considered porous medium and fluid characteristics are given 
in Table \ref{tab:value}. Fluid temperature is fixed to $T=303K$.

\begin{table}[htb] \centering
    \begin{tabular}{|c|rc||c|rc|}
        \hline
        \multicolumn{3}{|c||}{Porous medium parameters} & \multicolumn{3}{c|}{Fluid characteristics}\\
        \hline
        Parameter & \multicolumn{2}{|c||}{Value} & Parameter & \multicolumn{2}{c|}{Value} \\
        \hline
        $k$ & $5\;10^{-20}$ & $m^2$ & $D_l^h$ & $3\;10^{-9}$ & $m^2/s$ \\
        $\Phi$ & $0.15$ & $(-)$ &   $\mu_l$ & $1\;10^{-3}$ & $Pa.s$ \\
        $P_r$ & $2\;10^6$ & $Pa$ & $\mu_g$ & $9\;10^{-6}$ & $Pa.s$ \\
        $n$ & $1.49$ & $(-)$ & $H(T=303K)$ & $7.65\;10^{-6}$ & $mol/Pa/m^3$ \\
        $S_{lr}$ & $0.4$ & $(-)$ & $M_l$ & $10^{-2}$ & $kg/mol$ \\
        $S_{gr}$ & $0$ & $(-)$ & $M_g$ & $2\;10^{-3}$ & $kg/mol$ \\
        &&& $\rho_l^{std}$ & $10^3$ & $kg/m^3$ \\
        &&& $\rho_g^{std}$ & $8\;10^{-2}$ & $kg/m^3$ \\
        \hline
    \end{tabular}
    \caption{Values of porous medium parameters and fluid characteristics} \label{tab:value}
\end{table}

\subsubsection{Test case 1}
In the first test case we consider the domain
$\Omega^1=[0m\ ;\ 200m]\times[-10m\ ;\ 10m]$
with an impervious boundary
$\Gamma^1_{imp}=[0m\ ;\ 200m]\times\{-10m,10m\}$,
an inflow boundary
$\Gamma^1_{in}=\{0m\}\times[-10m\ ;\ 10m]$
and an outflow boundary
$\Gamma^1_{out}=\{200m\}\times[-10m\ ;\ 10m]$.
The following boundary conditions are imposed~:
\begin{itemize}
    \item $\phi_{tot}\cdot\nu=\phi^{h}\cdot\nu=0$ on the impervious boundary $\Gamma^1_{imp}$,
    \item $\phi_{tot}\cdot\nu=\phi^{h}\cdot\nu=Q_d^h$ on the inflow boundary $\Gamma^1_{in}$,
    \item $X=0$ and $p_l=p^1_{l,out}$ on the outflow boundary $\Gamma^1_{out}$
\end{itemize}
where $Q_d^h$ and $p^1_{l,out}$ are constant scalars.
Source terms are fixed to zero (${\cal F}^h={\cal F}^w=0$).
Initial conditions are $X(t=0)=0$ and $p_l(t=0)=p^1_{l,out}$ on $\Omega^1$.
The boundary parameters are fixed to $Q_d^h=1.5\:10^{-5}\;m/years$ and $p^1_{l,out}=10^6\;Pa$.

\subsubsection{Test case 2}
In the second test case we consider the domain $\Omega^2=[0m\ ;\ 200m]\times[-100m\ ;\ 100m]$
with an outflow boundary $\Gamma^2_{out}=\partial\Omega^2$
and where $B^2_h=[90m\ ;\ 110m]\times[-10m\ ;\ 10m]$ is the support of hydrogen source term.
On the outflow boundary $\Gamma^2_{out}$ we impose $X=0$ and $p_l=p^2_{l,out}$.
Source terms are defined by ${\cal F}^h=F^2_h\chi_{B^2_h}$ and ${\cal F}^w=0$.
Initial conditions are $X(t=0)=0$ and $p_l(t=0)=p^2_{l,out}$ on $\Omega^2$.
Here $p^2_{l,out}$ and $F^2_h$ are constant scalars fixed to $p^2_{l,out}=10^6\;Pa$ and $F^2_h=8\;10^{-13}\;kg/m^3/s\approx2.5\;10^{-5}\;kg/m^3/year$.

\subsection{Numerical results}
System (\ref{eq:1-3-fin}), (\ref{eq:2-3-fin}) is a coupled nonlinear partial differential equation system.
Numerical simulations
use an implicit scheme for time discretization, a finite volume scheme (using Multi Point Flux Approximation)
for space discretization
and a Newton-Raphson like method to solve nonlinearities. All computations are performed with the Cast3m
software (see \cite{castem}).

In both cases we present, at several times, spatial evolutions of the liquid pressure, the total hydrogen
molar density and the
 gas saturation along the line ${\cal L}_{cut}=[0m\ ;\ 200m]\times\{0m\}$. Computations are 
 performed since the time $t=0$ up to the stationary state.

\subsubsection{Results and comments}

\begin{figure}[hbtp] \centering
    \includegraphics[width=.7\textwidth]{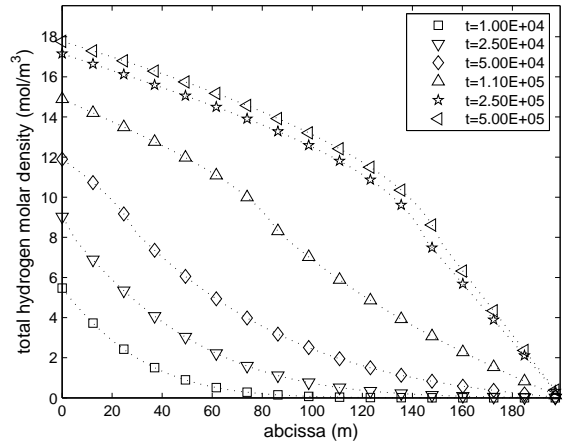}
    \caption{Test case 1 : spatial evolution along the line ${\cal L}_{cut}$ of the total hydrogen molar density $\frac{\rho_g^{std}}{M^h}X$ at several times $t$ (in years)}
    \label{fig:C1_H2}
\end{figure}

\begin{figure}[hbtp] \centering
    \includegraphics[width=.7\textwidth]{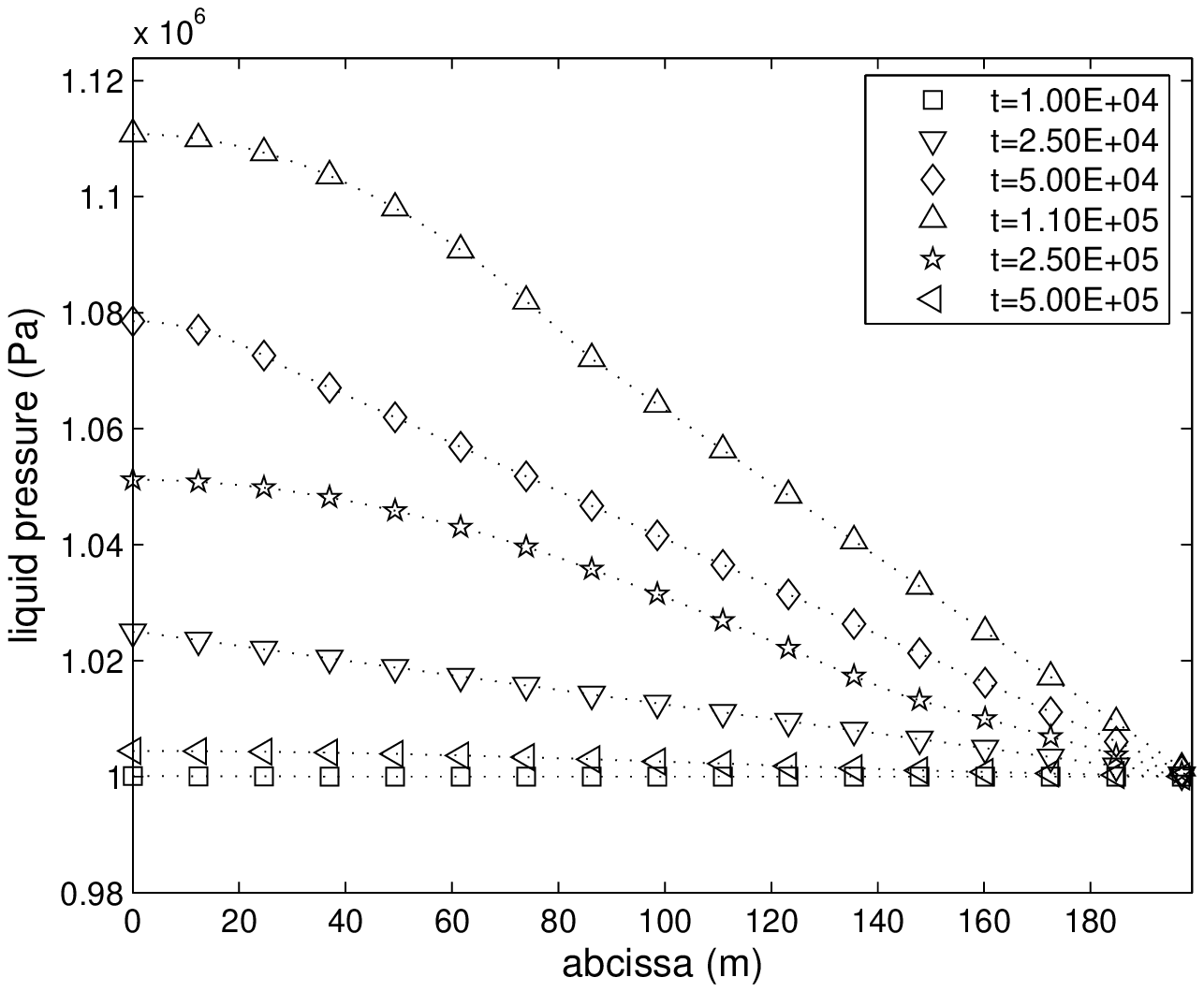}
    \caption{Test case 1 : spatial evolution along the line ${\cal L}_{cut}$ of the liquid pressure $p_l$ at several times $t$ (in years)}
    \label{fig:C1_pl}
\end{figure}

\begin{figure}[hbtp] \centering
    \includegraphics[width=.7\textwidth]{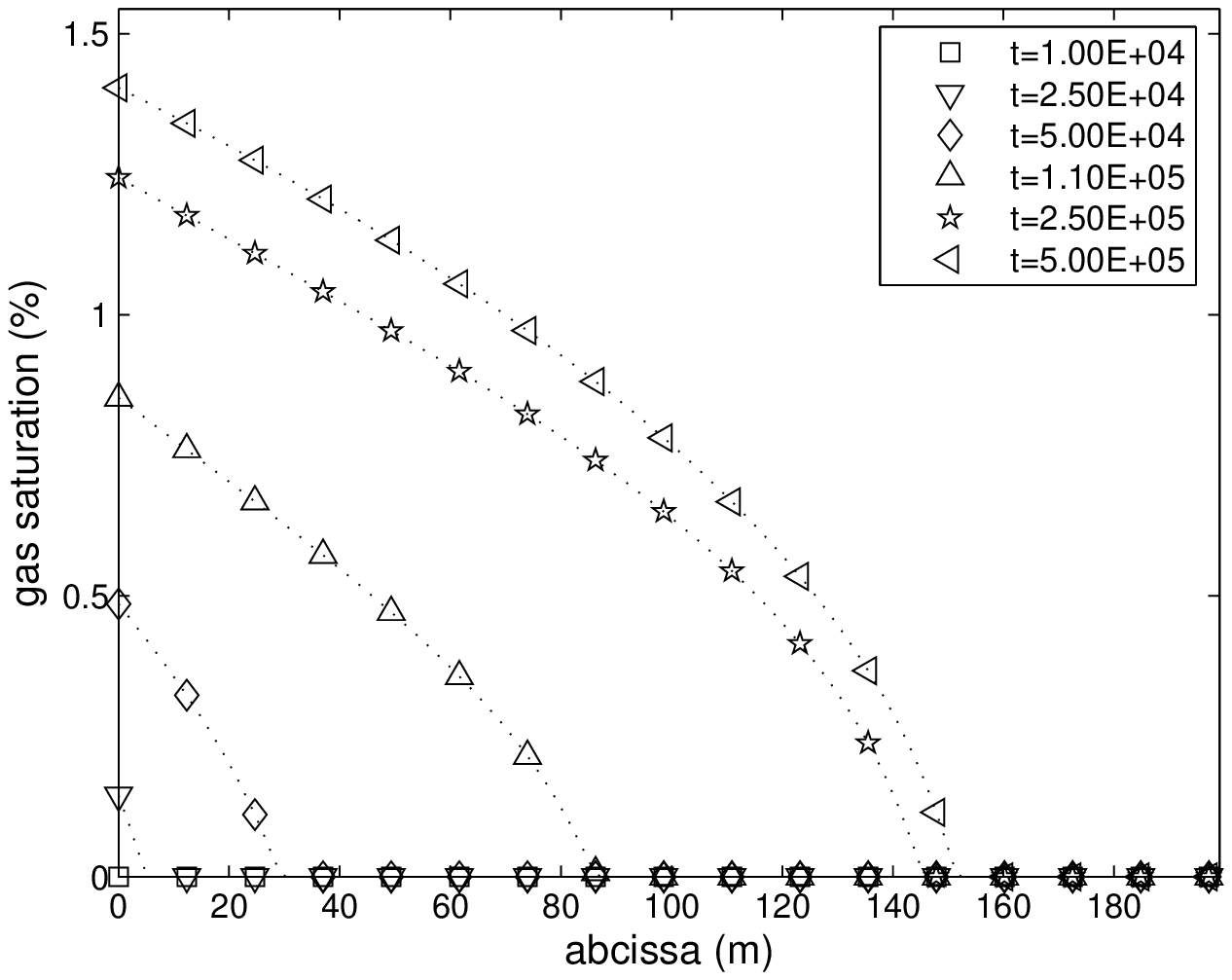}
    \caption{Test case 1 : spatial evolution along the line ${\cal L}_{cut}$ of the gas saturation $S_g$ at several times $t$ (in years)}
    \label{fig:C1_Sg}
\end{figure}

\begin{figure}[hbtp] \centering
    \includegraphics[width=.7\textwidth]{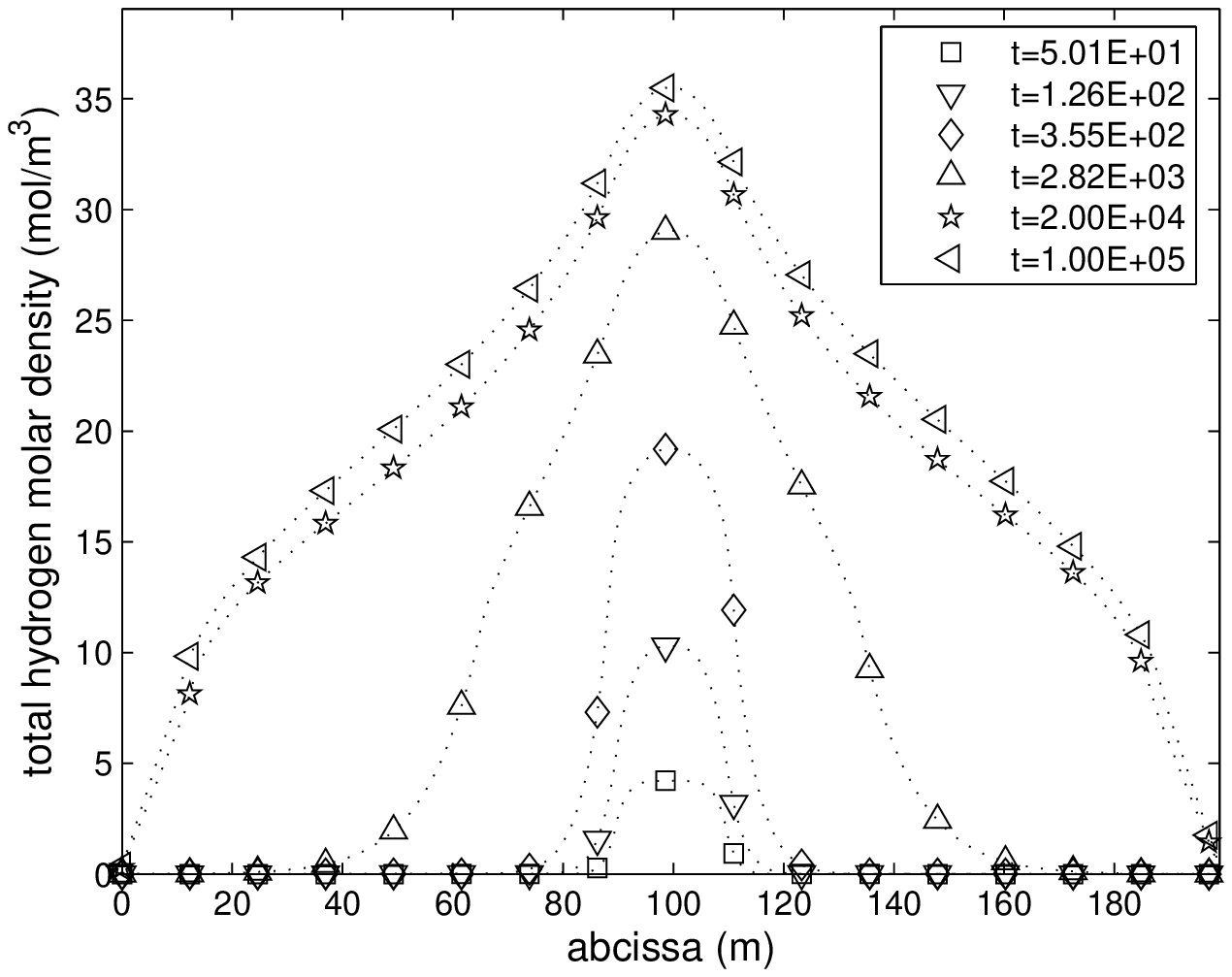}
    \caption{Test case 2 : spatial evolution along the line ${\cal L}_{cut}$ of the total hydrogen molar density $\frac{\rho_g^{std}}{M^h}X$ at several times $t$ (in years)}
    \label{fig:C2_H2}
\end{figure}

\begin{figure}[hbtp] \centering
    \includegraphics[width=.7\textwidth]{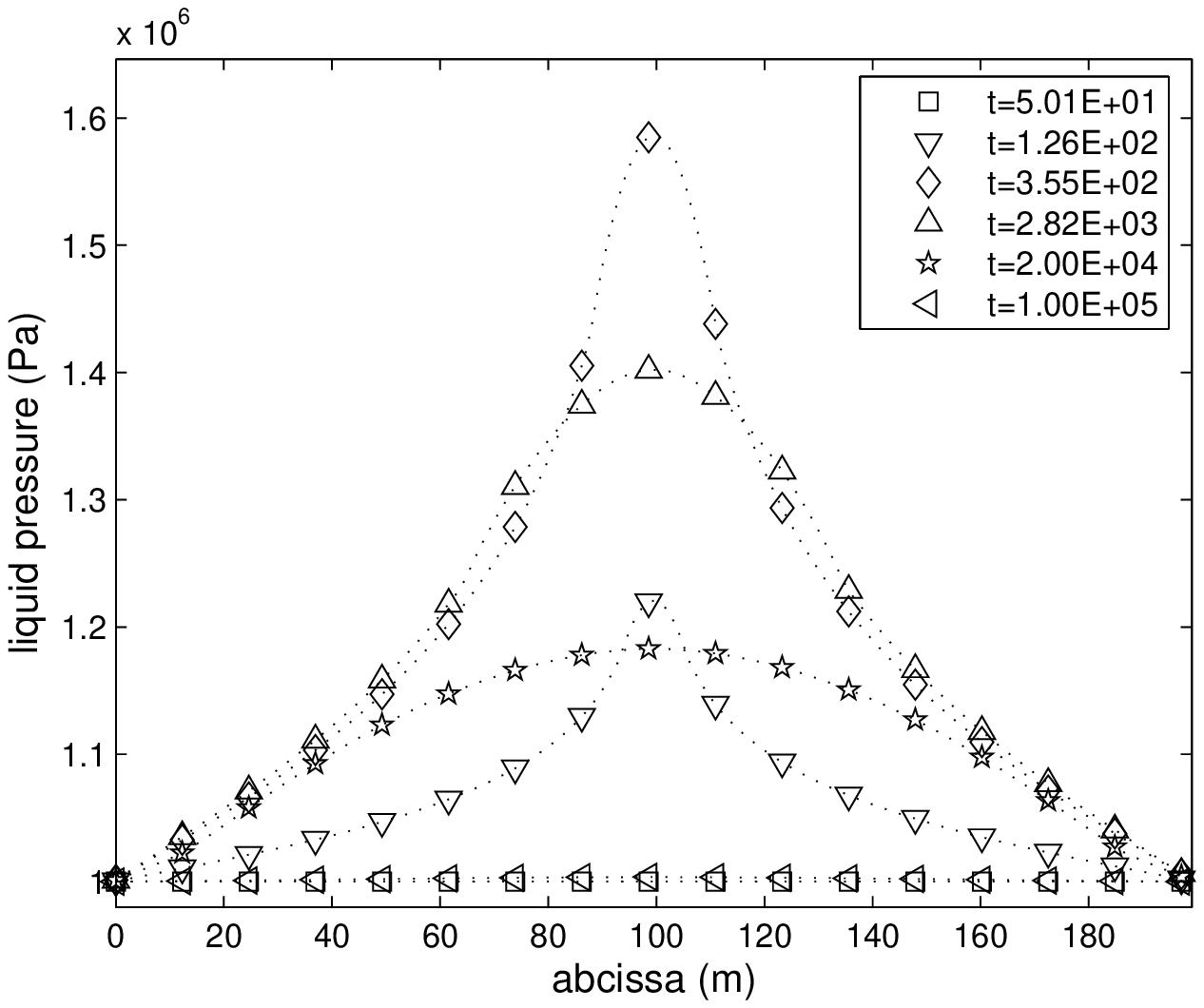}
    \caption{Test case 2 : spatial evolution along the line ${\cal L}_{cut}$ of the liquid pressure $p_l$ at several times $t$ (in years)}
    \label{fig:C2_pl}
\end{figure}

\begin{figure}[hbtp] \centering
    \includegraphics[width=.7\textwidth]{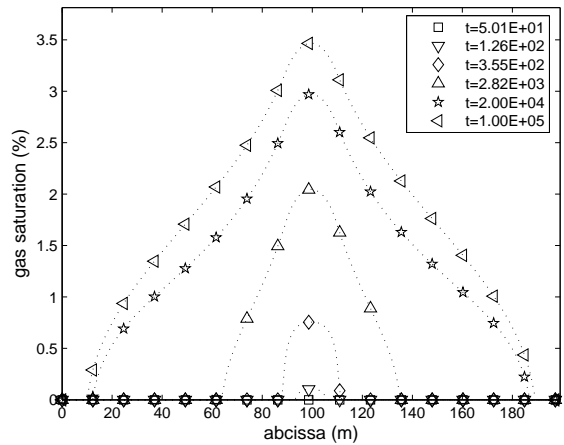}
    \caption{Test case 2 : spatial evolution along the line ${\cal L}_{cut}$ of the gas saturation $S_g$ at several times $t$ (in years)}
    \label{fig:C2_Sg}
\end{figure}

Results of test case 1 are plotted on figures \ref{fig:C1_H2}, \ref{fig:C1_pl} and \ref{fig:C1_Sg}. 
The \textbf{total hydrogen molar density}
$\frac{\rho_g^{std}}{M^h}X$ (Fig.\ref{fig:C1_H2}), the \textbf{liquid pressure} $p_l$ (Fig.\ref{fig:C1_pl})
and the \textbf{gas saturation} $S_g$
(Fig.\ref{fig:C1_Sg}) are plotted at times 
$t=1\;10^4$, $ 2.5\;10^4$, $5\;10^4$, $1.1\;10^5$, $2.5\;10^5$ and $5\;10^5\;years$.
Results of test case 2 are plotted on figures \ref{fig:C2_H2}, \ref{fig:C2_pl} and \ref{fig:C2_Sg}. 
The total hydrogen molar density
$\frac{\rho_g^{std}}{M^h}X$ (Fig.\ref{fig:C2_H2}), the liquid pressure $p_l$ (Fig.\ref{fig:C2_pl}) 
and the gas saturation $S_g$ (Fig.\ref{fig:C2_Sg}) are plotted at times 
$t=50.1$, $125$, $355$, $2820$, $2\;10^4$ and $10^5\;years$.

In both cases, we can identify three characteristic times : at
$t=T_1$ the gas phase appears; at $t=T_2$ the maximum liquid
pressure is reached; at $t=T_3$ the system is close to the
stationary state. For test case 1, we have $$T_1\approx
2\;10^4\;years,\ T_2\approx 1.1\;10^5\;years\ \mbox{and}\ T_3\approx
5\;10^5\;years;$$ for test case 2, we have $$T_1\approx 90\;years,\
T_2\approx 355\;years\ \mbox{and}\ T_3\approx 10^5\;years.$$ Global
behaviors of both cases are similar and can be summarized as
follow~:
\begin{itemize}
\item For $0\leq t<T_1$~: only total hydrogen density increases while liquid pressure and gas saturation stay constant; 
    during this stage $X<C_hp_l$ and all the domain is saturated in water ($S_g=0$).
\item From $t=T_1$, $X\geq C_hp_l$ in a part of the domain meaninig that gas phase exists ($S_g>0$) in this part.
\item For $T_1\leq t\leq T_2$~: while gas phase appears, liquid pressure increases and a non zero pressure 
    gradient appears what corresponds to a fluid displacement according to the Darcy-Muskat law. 
    Total hydrogen density and gas saturation increase and the unsaturated area grows.
\item For $T_2\leq t$~: while total hydrogen density and gas saturation continue to increase, 
    liquid pressure and pressure gradient decrease. When $t\rightarrow\infty$, the system reach a 
    stationary state where saturated and unsaturated areas coexist and liquid pressure gradient is null.
\end{itemize}



\section{Concluding remarks}
From balance equations, constitutive relations and equations of
state, assuming thermodynamical equilibrium, we have derived a model
for describing underground gas migration in  water saturated or
unsaturated porous media, including diffusion of components in
phases and capillary effects. In the last part of this paper,
numerical simulations on simplified situations inspired by the
"Couplex-gas" benchmark \cite {Bench}, show evidence of its ability
: - to describe gas (hydrogen) generation and migration - and to
treat
 the difficult problem, as it appeared in the results of
"Couplex-gas" \cite {Bench}, of correctly simulating evolution of
the unsaturated region, in a deep
geological repository, created by gas generation. A forthcoming paper will be devoted to the use of this model for solving the
"Couplex-gas" benchmark \cite {Bench} and other 3-D situations of gas migration in  water saturated or
unsaturated porous media, including the  design a of
numerical test cases synthesizing the main challenges appearing
in gas generation and migration.


\begin{acknowledgements}
This work was partially supported by GdR
MoMaS (PACEN/CNRS, ANDRA, BRGM, CEA, EDF, IRSN). Most of the work on this paper was done when
Mladen Jurak was visiting Universit\'{e} Lyon 1; we thank CNRS, UMR 5208, Institut Camille Jordan, for
hospitality.
\end{acknowledgements}


\end{document}